\documentclass[draftclsnofoot,onecolumn,12pt]{IEEEtran}
\usepackage{multirow}
\usepackage{booktabs}
\usepackage{amsfonts}
\usepackage{amsmath,cases}
\usepackage{amssymb}
\usepackage{graphicx}
\usepackage{cite}
\usepackage{epsfig}
\usepackage{url}
\usepackage{algorithm}
\usepackage{algorithmic}
\usepackage{epstopdf}
\usepackage{color}
\usepackage[tight]{subfigure}

\newcommand{\ds}{\displaystyle}
\newtheorem{myth}{Theorem}
\newtheorem{mypro}{Proposition}

\newcommand{\la}{\langle}
\newcommand{\ra}{\rangle}

\newcommand{\qed}{\mbox{}\hfill$\Box$
\vskip 3mm}
\newcommand{\Prf}{{\it Proof: \ }}

\newcommand{\bt}{\pmb{t}}

\newcommand{\Prob}{\mbox{Prob}}
\newcommand{\CN}{{\cal CN}}
\newcommand{\bw}{\mathbf{w}}

\newcommand{\br}{\mathbf{r}}

\newcommand{\rik}{r_i^{(\kappa)}}

\newcommand{\wjk}{w_j^{(\kappa)}}
\newcommand{\wik}{w_i^{(\kappa)}}
\newcommand{\wk}{w^{(\kappa)}}

\newcommand{\bR}{\mathbf{R}}
\newcommand{\xik}{x_i^{(\kappa)}}
\newcommand{\xeik}{x_{e,i}^{(\kappa)}}

\newcommand{\bz}{\mathbf{z}}

\newcommand{\wmink}{w_{\min}^{(\kappa)}}

\begin{document}
\title{Optimal Beamforming for Physical Layer Security in MISO Wireless Networks}
\author{Z. Sheng, H. D. Tuan,  T. Q. Duong and H. V.  Poor
\thanks{Zhichao Sheng and Hoang Duong Tuan are with the School of Electrical and Data Engineering, University of Technology Sydney, Broadway, NSW 2007, Australia (email: zhichaosheng@163.com, Tuan.Hoang@uts.edu.au)}
\thanks{Trung Q. Duong is with Queen's University Belfast, Belfast BT7 1NN, UK  (email: trung.q.duong@qub.ac.uk)}
\thanks{H. Vincent Poor is with the Department of Electrical Engineering, Princeton University, Princeton, NJ 08544, USA (e-mail: poor@princeton.edu)} }
\date{}
\maketitle
\vspace*{-1.8cm}
\begin{abstract}
A wireless network of multiple transmitter-user pairs overheard by an eavesdropper,
where the transmitters are equipped with multiple
antennas while the users and eavesdropper are equipped with a single antenna, is considered.
At different levels of wireless channel knowledge, the problem of interest is beamforming
to optimize the users' quality-of-service (QoS)
in terms of their secrecy throughputs or maximize the network's energy efficiency under users' QoS. All these problems
are seen as very difficult optimization problems with many nonconvex constraints and nonlinear equality constraints in
beamforming vectors. The paper develops path-following computational procedures of low-complexity and rapid convergence
for the optimal beamforming solution. Their practicability is demonstrated through numerical examples.
\end{abstract}
\vspace*{-0.5cm}


\section{INTRODUCTION} \label{sec:Intro}
Securing information has emerged  as one of the most critical issues in wireless communication \cite{FTA13,MFHS14}. The broadcast nature
of wireless transmissions implies that  they can be quite vulnerable to adversary, who attempts to intercept
their information delivery or overhear the confidential information intended for their users \cite{LPS08,Baetal13}.
Physical layer security (PLS) exploiting the physical properties of  wireless channels
\cite{P12,PS17} has been proposed to ensure the secrecy of data transmissions to end-users
of low complexity, for which encryption cannot be used. PLS is based
on information theoretic characterizations of secrecy, under which  the user secrecy throughput
of a wireless transmission overheard by eavesdroppers is determined as
the difference between the user throughput and eavesdroppers' throughput \cite{Cetal13,MFHS14}. Transmit
beamforming to improve the user throughput
while controlling the throughput of the wiretapped signal at the eavesdroppers thus presents an effective way for secrecy
throughput enhancement. Beamforming design for maximizing instantaneous secrecy  throughput  has been considered  in \cite{MS11,LM13,LZC14,Zhaetal15,Chuetal15} by semi-definite relaxation and randomization with the known inefficiency
\cite{PTKD12}. This beamforming design has been successfully addressed in \cite{Naetal17a,Naetal17b}.
In regards to outage probability, several works such as \cite{PSKP12,Zhuetal16} used the Bernstein-type inequalities
obtained in an unpublished work \cite{Be09}. We will show that the results based on such Bernstein-type inequalities may be
very conservative. Reference \cite{Staetal15} considered outage region characterization of given beamformers under imperfect channel state information (CSI).

On the other hand, as  energy efficiency (EE) became  a very serious concern
in wireless communication \cite{Caetal14,Ietal14},
the secure energy efficiency (SEE), which is the ratio of the secrecy throughput to the total network power consumption, measured in terms of secrecy bits per Joule per Hertz is also increasingly  important in SPL \cite{WBCH16,NTDP17}. Exploiting
the perfect CSI, the SEE maximization in \cite{ZYS15,VKT16,Vuetal16}
is based on costly  beamformers, which completely cancel
the multi-user interference and wiretapped signal at the eavesdroppers.  The computational complexity of
the SEE optimization algorithms for single-user multi-input multi-output (MIMO)/single-input single-output (SISO) communications in \cite{Kaetal15} and \cite{ZLJ16} is also high as each iteration still involves a difficult
nonconvex optimization problem. Our previous work \cite{NTDP17} considered SEE optimization for a more general case of MIMO networks. SEE optimization was also considered in \cite{Naetal17b} for the worst case of uncertainties for users' and eavesdroppers'
channels. There is no existing work on SEE optimization with secrecy throughput in terms of probability outage.

In this paper, we consider a network of multiple transmitter-user pair overheard by an eavesdropper. As the transmitters
are assumed to be equipped with multiple antennas while the users and eavesdropper are equipped with a single antennas,
the target is to design transmit beamformers to optimize either the users' quality-of-service (QoS) in terms of their
secrecy throughput or the network's SEE under  the users' QoS. It should be realized that these problems of beamforming
design are still widely open for research, so we consider them at different levels of channel knowledge. The paper
is structured as follows.  Section II is devoted to the problem statements. Section III considers these problems under
the perfect CSI of the all concerned channels, where path-following algorithms of low complexity are developed for their
solution. In Section IV, the perfect CSI of the channels between the transmitters and user is assumed but only
the distribution of the channels between the transmitters and eavesdropper is assumed known. As such,
the eavesdropper's throughput is not deterministically defined but is defined through its probability outage, which
leads to a nonlinear equation in beamforming vectors and the eavesdropper's throughput, making the beamforming designs
much more computationally challenging. Under the same knowledge on the channels between the transmitters and
eavesdropper in Section IV, Section V also assumes that the channels between the transmitters and users are uncertain
with Gaussian distributed errors, under which there is no known result on the probability outage of the users' throughput.
Nevertheless, based on a new result on outage probability obtained in Appendix I, both problems of users' QoS optimization
and network's SEE optimization are successfully addressed.  The simulation Section V shows
the efficiency of the path-following algorithms developed in sections III-V. Conclusions are given in Section VI.
 Appendix I provides a new result on both
upper bound and lower bound of the outage-aware user throughput. Appendix II shows the conservativeness of some
other results, which are based on Bernstein type inequalities. Some fundamental deterministic inequalities that are used
in Sections III-V are given in Appendix III.

\emph{Notation.} The inner product between vectors $x$ and
$y$ is defined as $\langle x,y\rangle=x^Hy$. Analogously, $\la
X,Y\ra={\sf Trace}(X^HY)$ for matrices
$X$ and $Y$. Optimization variables are boldfaced. Also the notation  $\sum_{j\neq i}^M$ refers
to the summation taking over the index set $\{1,\dots, M\}\setminus \{i\}$. $I$ is the identity matrix
of appropriate dimension and ${\cal CN}(0,I)$ is the set of complex Gaussian random variables of zero means
and identity covariance.
\section{Problem statements}
Consider a communication network of  $M$ transmitter-user pairs overheard by an eavesdropper.
Each transmitter is equipped with  $N_t$ transmit antennas while the users and eavesdropper
are equipped by a single antenna. Thus, without the eavesdropper, the network looks very much like that
considered in \cite{AV11,KTN11,LCLC13,LCC15b,Cheetal15}. Each information $s_i$ for user $i$, which is normalized
 to $E(s_i^2)=1$, is beamformed by $\bw_i\in \mathbb{C}^{N_t}$. The received signal at user $i$ is
\begin{equation}\label{sec1}
y_i=h_{ii}^H\bw_is_i+\ds\sum_{j\neq i}^Mh_{ji}^H\bw_js_j + n_i,
\end{equation}
where $h_{ji}\in\mathbb{C}^{N_t}$ is the vector channel  from transmitter $j$ to user $i$ and
$n_i$ is the background noise with power $\sigma_i^2$.

Analogously, the received signal at the eavesdropper is
\begin{equation}\label{sec2}
y_E=\sum_{i=1}^Mh_{ie}^H\bw_ix_i+ n_e,
\end{equation}
where $h_{ie}\in\mathbb{C}^{N_t}$  is the vector channel from transmitter $i$ to the eavesdropper  and
$n_e$ is the background noise with power $\sigma_e^2$.

For $\bw\triangleq [\bw_i]_{i=1,\dots,M}$, suppose that $f_i(\bw)$ is the throughput
user $i$ while $g_i(\bw)$ is the wiretapped throughput for user $i$ at the eavesdropper. Our interest is
the following optimization problems.
\begin{itemize}
\item Secrecy throughput maximin optimization under transmitters' power constraints:
\begin{subequations}\label{sec4}
\begin{eqnarray}
\max_{\pmb{w}}\ \Phi(\bw)\triangleq\min_{i=1,...,M}[f_i(\bw) -g_i(\bw)]\quad\mbox{s.t.}\label{sec4a}\\
||\bw_i||^2\leq P_i, i=1,\dots, M,\label{sec4b}
\end{eqnarray}
\end{subequations}
with  $P_i$ given to set the limit of transmission power at transmitter $i$.
\item Energy efficiency maximization over the secrecy throughput threshold constraints:
\begin{subequations}\label{esec4}
\begin{eqnarray}
\ds\max_{\pmb{w}}\ \Theta(\bw)\triangleq \left[\ds\sum_{i=1}^M[f_i(\bw)-g_i(\bw)]\right]/\pi(\bw) \quad\mbox{s.t.}\quad (\ref{sec4b}),\label{esec4a}\\
f_i(\bw)-g_i(\bw)\geq c_i,\ i=1,\dots, M,
\label{esec4b}
\end{eqnarray}
\end{subequations}
with given $c_i$ to set the QoS threshold for user $i$ and the total network power consumption $\pi(\bw)\triangleq \zeta\ds\sum_{i=1}^M||\bw_i||^2+P_c$   in transmitting $\bw_is_i$, where
$0<\zeta<1$ is the the reciprocal of the drain efficiency of the power amplifier and
$P_c=\sum_{i=1}^MP_c^i$ with circuit power $P_c^i$ at transmitter $i$.
\end{itemize}
\section{Instantaneous secrecy throughput optimization}
When the perfect CSI of all channels is available at the transmitters, the user $i$' instantaneous throughput
is defined by
\begin{equation}\label{userrate}
f_i(\bw)\triangleq \ds\ln\left(1+\frac{|h^H_{ii}\bw_i|^2}{\sum_{j\neq i}|h^H_{ji}\bw_j|^2+\sigma_i^2}\right),
\end{equation}
while the instantaneous wiretapped throughput  for user $i$ at the eavesdropper is defined by
\begin{equation}\label{eveas}
g_i(\bw)\triangleq \ds\ln\left(1+\frac{|h^H_{ie}\bw_i|^2}{\sum_{j\neq i}
|h^H_{je}\bw_j|^2+\sigma_e^2}\right).
\end{equation}
For $f_i(\bw)$ and $g_i(\bw)$ defined by (\ref{userrate}) and (\ref{eveas}),
problem (\ref{sec4}) is a particular case of the multi-cell beamforming design
that considered in \cite{Naetal17a,Naetal17b}. We now propose a more efficient computation tailored for (\ref{sec4}).

Let   $w^{(\kappa)}$ be a feasible point for (\ref{sec4}) found from $(\kappa-1)$th iteration.
Applying  inequality (\ref{sec5.2}) in the Appendix II
for $x=1/|h^H_{ii}\bw_i|^2$, $y=\sum_{j\neq i}^M|h^H_{ji}\bw_j|^2+\sigma_i^2$,
and $\bar{x}=1/|h^H_{ii}\wik|^2$, $\bar{y}=\sum_{j\neq i}^M|h^H_{ji}\wjk|^2+\sigma_i^2$
yields
\begin{eqnarray}
f_i(\bw)\geq f_i^{(\kappa)}(\bw)\label{dc1}
\end{eqnarray}
for
\begin{eqnarray}
f_i^{(\kappa)}(\bw)&\triangleq&\ds\ln(1+\xik)+\ds\frac{\xik}{1+\xik}\left(2-
\frac{|h^H_{ii}\wik|^2}{{\color{black}2\Re\{(\wik)^Hh_{ii}h^H_{ii}\bw_i\}}-|h^H_{ii}\wik|^2}\right.\nonumber\\
&&\left.
-
\ds\frac{\sum_{j\neq i}^M|h^H_{ji}\bw_j|^2+\sigma_i^2}{\sum_{j\neq i}^M|h^H_{ji}\wjk|^2+\sigma_i^2}\right),\label{dc3}
\end{eqnarray}
over the trust region
\begin{equation}\label{trustw}
{\color{black}2\Re\{(\wik)^Hh_{ii}h^H_{ii}\bw_i\}}-|h^H_{ii}\wik|^2>0, i=1,\dots, M,
\end{equation}
where
\[
\xik=\frac{|h^H_{ii}\wik|^2}{\sum_{j\neq i}^M|h^H_{ji}\wjk|^2+\sigma_i^2}.
\]
On the other hand, applying inequality (\ref{conv1}) in the Appendix II for
$x=|h^H_{ie}\bw_i|^2$, $y=\sum_{j\neq i}^M|h^H_{je}\bw_j|^2{\color{black}+\sigma_e^2}$
and $ \bar{x}=|h^H_{ie}\wik|^2$, $\bar{y}=\sum_{j\neq i}^M|h^H_{je}\wjk|^2{\color{black}+\sigma_e^2}$
yields
\begin{equation}\label{dc2}
g_i(\bw)\leq g_i^{(\kappa)}(\bw),
\end{equation}
for
\begin{eqnarray}
g_i^{(\kappa)}(\pmb{w})&\triangleq&\nonumber\\
\ln(1+\xeik)+\ds\frac{1}{1+\xeik}\left(\frac{|h^H_{ie}\bw_i|^2}{\sum_{j\neq i}^M(
2\Re\{(\wjk)^Hh_{je}h^H_{je}\bw_j\}-|h^H_{je}\wjk|^2)+\sigma_e^2}  -\xeik\right),&&
\label{dc4}
\end{eqnarray}
over the trust region
\begin{equation}\label{trustwe}
{\color{black}\sum_{j\neq i}^M\left(2\Re\{(\wjk)^Hh_{je}h^H_{je}\bw_j\}-|h^H_{je}\wjk|^2\right)}>0, i=1,\dots,M,
\end{equation}
where
\[
\xeik=|h^H_{ie}\wik|^2/(\sum_{j\neq i}^M
|h^H_{je}\wjk|^2+\sigma_e^2).
\]
At the $\kappa$-th iteration we solve the following convex optimization problem
to generate the next feasible point $w^{(\kappa+1)}$:
\begin{equation}\label{dc5}
\max_{\pmb{w}}\ \Phi^{(\kappa)}(\bw)\triangleq \min_{i=1,...,M} [f_i^{(\kappa)}(\pmb{w})-g_i^{(\kappa)}(\pmb{w})]\quad\mbox{s.t.}\quad (\ref{sec4b}),
{\color{black} (\ref{trustw}), (\ref{trustwe})}
\end{equation}
From (\ref{dc1}) and (\ref{dc2}), it can be easily checked that
$\Phi(\bw)\geq \Phi^{(\kappa)}(\bw)\ \forall\ \bw$
and $\Phi(w^{(\kappa)})= \Phi^{(\kappa)}(w^{(\kappa)})$. On the other hand,
$\Phi^{(\kappa)}(w^{(\kappa+1)})> \Phi^{(\kappa)}(\wk)$ as far as $w^{(\kappa+1)}\neq w^{(\kappa)}$
because the former is the optimal solution of (\ref{dc5}) while the latter is a feasible point for
(\ref{dc5}). We thus have the following chain of inequalities and equalities:
\[
\Phi(w^{(\kappa+1)})\geq \Phi^{(\kappa)}(w^{(\kappa+1)})> \Phi^{(\kappa)}(\wk)=\Phi(\wk),
\]
which implies that $w^{(\kappa+1)}$ is a better feasible point than $w^{(\kappa)}$
for the nonconvex optimization problem (\ref{sec4}). Using a similar  convergence argument as
 \cite{MW78}, we can show that at least the sequence $\{w^{(\kappa)}\}$ converges to its locally optimal
 solution.  As such, the proposed  Algorithm \ref{alg1} a path-following computational procedure for (\ref{dc4}).
\begin{algorithm}
\caption{Path-following algorithm for maximin instantaneous secrecy throughput optimization} \label{alg1}
\begin{algorithmic}
\STATE \textbf{Initialization}: Set $\kappa=0$. Choose an initial feasible point
$w^{(0)}$ for the convex constraints (\ref{sec4b}).
Calculate $R_{\min}^{(0)}$ as the value of the objective in (\ref{sec4}) at $w^{(0)}$. Set $\kappa=0$.
\REPEAT
\STATE $\bullet$ Solve the
convex optimization problem \eqref{dc5} to obtain the
solution $w^{(\kappa+1)}$.
\STATE $\bullet$ Calculate $R_{\min}^{(\kappa+1)}$ as the value of the objective in (\ref{sec4}) at $w^{(\kappa+1)}$.
 \STATE $\bullet$ Reset
$\kappa+1\rightarrow \kappa$.
 \UNTIL{$\frac{R_{\min}^{(\kappa+1)}-R_{\min}^{(\kappa)})}{R_{\min}^{(\kappa)}} \leq
 \epsilon_{\rm tol}$}.
\end{algorithmic}
\end{algorithm}

Next, we address the EE maximization (\ref{esec4}). A direct approach (see e.g. \cite{Naetal17b})
is based on a lower bounding approximation for the objective function in (\ref{esec4a}).
We now propose another approach, which uses the above approximation for the numerator of the objective function only,
so the EE maximization problem (\ref{esec4}) is indeed not more computationally difficult than
the throughput optimization problem (\ref{sec4}).

As before, let $w^{(\kappa)}$ be its feasible point found from $(\kappa-1)$th iteration. At the $\kappa$-th iteration, we solve the following convex optimization problem to generate the next feasible point $w^{(\kappa+1)}$:
\begin{eqnarray}\label{eseck1}
\max_{\pmb{w}}\sum_{i=1}^{M} [f_i^{(\kappa)}(\pmb{w})-g_i^{(\kappa)}(\pmb{w})]
-\Theta(\wk)\pi(\bw)\quad\mbox{s.t.}\quad (\ref{sec4b}),
{\color{black} (\ref{trustw}), (\ref{trustwe})},\nonumber\\
f_i^{(\kappa)}(\pmb{w})-g_i^{(\kappa)}(\pmb{w})\geq c_i, i=1,\dots, M.
\end{eqnarray}
Note that $w^{(\kappa)}$ is a feasible point for (\ref{eseck1}), under which
\[
\sum_{i=1}^{M} [f_i^{(\kappa)}(\wk)-g_i^{(\kappa)}(\wk)]
-\Theta(\wk)\pi(\wk)=0.
\]
Therefore, as far as $w^{(\kappa+1)}\neq \wk$, the optimal solution $w^{(\kappa+1)}$ of (\ref{eseck1}) must satisfy
\[
\sum_{i=1}^{M} [f_i^{(\kappa)}(\wk)-g_i^{(\kappa)}(\wk)]
-\Theta(\wk)\pi(w^{(\kappa+1)})>0,
\]
i.e.
so
\[
\begin{array}{lll}
\Theta(w^{(\kappa+1)})&\triangleq&
\ds\sum_{i=1}^M[f_i(w^{(\kappa+1)})-g_i(w^{(\kappa+1)})]/\pi(w^{(\kappa+1)})\\
&\geq&\ds\sum_{i=1}^M[f_i^{(\kappa)}(w^{(\kappa+1)})-g_i^{(\kappa)}(w^{(\kappa+1)})]/\pi(w^{(\kappa+1)})\\
&>&\Theta(\wk),
\end{array}
\]
implying that $w^{(\kappa+1)}$ is a better feasible point than $\wk$ for the nonconvex optimization problem (\ref{esec4}).
As such, Algorithm \ref{alg1e}, which is different from Algorithm \ref{alg1} by solving the convex optimization
problem \eqref{eseck1} at the $\kappa$-th iteration to generate the next feasible point $w^{(\kappa+1)}$ instead of (\ref{dc5}) in Algorithm \ref{alg1}, at least converges to a locally optimal solution.\\
A feasible point $w^{(0)}$ for (\ref{esec4}) in the initialization of Algorithm \ref{alg1e}
is found by using Algorithm \ref{alg1}) in solving the problem
\begin{equation}\label{eefeas}
\max_{\bw} \min_{i=1,\dots, M} [f_i(\bw)-g_i(\bw_i)]/c_i\quad\mbox{s.t.}\quad (\ref{sec4b}).
\end{equation}
Namely Algorithm \ref{alg1} will terminate whenever
\[
\min_{i=1,\dots, M}[f_i(\wk)-g_i(\wk)]/c_i\geq 1.
\]
\begin{algorithm}
\caption{Path-following algorithm for EE optimization} \label{alg1e}
\begin{algorithmic}
\STATE \textbf{Initialization}: Set $\kappa=0$. Choose an initial feasible point
$w^{(0)}$ for (\ref{esec4}). Set $\kappa=0$.
\REPEAT
\STATE $\bullet$ Solve the
convex optimization problem \eqref{eseck1} to obtain the
solution $w^{(\kappa+1)}$.
 \STATE $\bullet$ Reset
$\kappa+1\rightarrow \kappa$.
 \UNTIL{$\frac{\Theta(w^{(\kappa+1)})-\Theta(w^{(\kappa)})}{\Theta(w^{(\kappa)})} \leq
 \epsilon_{\rm tol}$}.
\end{algorithmic}
\end{algorithm}
\section{Eavesdropper's outage probability maximization}
When the eavesdropper is no longer  part of the legitimate network, the assumption on the perfect CSI
for the wiretapped channels $h_{je}$ at the transmitters made in the previous section is not practical.
Instead, it is common to assume that only the wiretapped channel distribution
\begin{equation}\label{eve}
h_{je}=\sqrt{\bar{h}_{je}}\chi_{j},
\chi_j\in\CN(0,I), j=1,\dots, M
\end{equation}
is known, where $\sqrt{\bar{h}_{je}}$ is a deterministic quantity which is usually
dependent on the distance from the transmitter $j$ to the eavesdropper.
The user throughput $f_i(\bw)$ is still defined by (\ref{userrate}) but the
wiretapped throughput $g_i(\bw)$ for user $i$ at the eavesdropper  is defined via the following outage probability instead of the instantaneous throughput defined by (\ref{eveas}):
\begin{equation}\label{eveas1}
\max\{\ \ln(1+\br_i)\ :\ \Prob\left(\frac{\bar{h}_{ie}|\chi_i^H\bw_i|^2}
{\sum_{j\neq i}^M\bar{h}_{je}|\chi_j^H\bw_j|^2+\sigma_e^2}<\br_i\right)<\epsilon_{EV}\}
\end{equation}
for $\epsilon_{EV}>0$.
Note that $|\chi_j^H\bw_j|^2$ is an exponential distribution with mean $||\bw_j||^2$. Therefore,
by \cite{Shetal17}, this throughput is $\ln(1+\br_i)$, where
\begin{equation}\label{eveas2}
g_{i,o}(\bw,\br_i)= 0
\end{equation}
for
\begin{equation}\label{eveas1d}
g_{i,o}(\bw,\br_i)\triangleq \bar{h}_{ie}\ln(1-\epsilon_{EV})+\sigma_e^2\frac{\br_{i}}{||\bw_i||^2}+
\bar{h}_{ie}\sum_{j\neq i}^M\ln\left(1+\frac{\br_{i}\bar{h}_{je}||\bw_j||^2}{\bar{h}_{ie}||\bw_i||^2}\right),
\end{equation}
which increases in $\br_i$ with $\bw$ held fixed.\\
Similarly to \cite[Prop. 1]{Shetal17} the problem of secrecy rate maximin optimization (\ref{sec4})
is equivalently formulated by
\begin{subequations}\label{sec5}
\begin{eqnarray}
\max_{\pmb{w},\br}\min_{i=1,...,M} [\ln(1+\ds\frac{|h^H_{ii}\bw_i|^2}{\sum_{j\neq i}|h^H_{ji}\bw_j|^2+\sigma_i^2})-\ln(1+\br_i)]\quad\mbox{s.t}\quad (\ref{sec4b})\label{sec5a}\\
g_{i,o}(\bw,\br_i)\geq 0, \ i=1,...,M,
\label{sec5b} \\
{\color{black}\br_{i}>0}, \label{sec5c}
\end{eqnarray}
\end{subequations}
where the nonlinear equality constraint in (\ref{eveas1}) has been replaced by the nonconvex constraint (\ref{sec5b}).

The main difficulty is to develop a lower bounding approximation for the function $g_{i,o}(\bw,\br_i)$
at a feasible point $(\wk,r^{(\kappa)})$ for (\ref{sec5}), which is found from $(\kappa-1)$th iteration.
 Applying inequality (\ref{sec5.2}) for $x=1/\br_{i}\bar{h}_{je}||\bw_j||^2$,
$y=\bar{h}_{ie}||\bw_i||^2$, and $\bar{x}=1/\rik \bar{h}_{je}||\wjk||^2$, $\bar{y}=\bar{h}_{ie}||\wik||^2$ yields
\begin{equation}\label{dc7}
\ln\left(1+\frac{\br_{i}\bar{h}_{je}||\bw_j||^2}{\bar{h}_{ie}||\bw_i||^2}\right)\geq
\lambda_{ij}^{(\kappa)}(\br_{i},\bw_j,\bw_i)
\end{equation}
over the trust region
\begin{equation}\label{trust1}
2\Re\{(w_j^{(\kappa)} )^H\bw_j \}-||w_j^{(\kappa)}||^2>0
\end{equation}
for
\begin{eqnarray}\label{dc8}
\lambda_{ij}^{(\kappa)}(\br_{i},\bw_j,\bw_i)&\triangleq& \ln(1+x_{ij}^{(\kappa)})+y_{ij}^{(\kappa)}
(2-\frac{\rik\bar{h}_{je}||\wjk||^2}{\br_{i}\bar{h}_{je}(2\Re\{(w_j^{(\kappa)} )^H\bw_j \}-||w_j^{(\kappa)}||^2)}-\frac{\bar{h}_{ie}||\bw_i||^2}{\bar{h}_{ie}||\wik||^2} ) \nonumber \\
&=& \ln(1+x_{ij}^{(\kappa)})+y_{ij}^{(\kappa)}
(2-\frac{\rik ||\wjk||^2}{\br_{i}(2\Re\{(w_j^{(\kappa)} )^H\bw_j \}-||w_j^{(\kappa)}||^2)}-\frac{||\bw_i||^2}{||\wik||^2} )
\end{eqnarray}
 and $x_{ij}^{(\kappa)}\triangleq \rik\bar{h}_{je}||\wjk||^2 / \bar{h}_{ie}||\wik||^2$ and $y_{ij}^{(\kappa)}\triangleq x_{ij}^{(\kappa)}/(x_{ij}^{(\kappa)}+1)$.

Furthermore, applying inequality (\ref{conv3}) in the Appendix yields
\begin{equation}\label{dc9}
\frac{\br_{i}}{||\bw_i||^2}\geq \beta_i^{(\kappa)}(\br_i,\bw_i)
\end{equation}
where
\begin{equation}\label{dc10}
\beta_i^{(\kappa)}(\br_i,\bw_i)\triangleq 2\left(\sqrt{\rik}/||\wik||^2\right)\sqrt{\br_i}-\left(\rik/||\wik||^4\right)||\bw_i||^2,
\end{equation}
which is a concave function.

Based on (\ref{dc7}) and (\ref{dc9}) we obtain
\begin{equation}\label{add1}
g_{i,o}(\bw,\br_i)\geq g_{i,o}^{(\kappa)}(\bw,\br_i)
\end{equation}
for
\begin{equation}\label{add2}
g_{i,o}^{(\kappa)}(\bw,\br_i)\triangleq \bar{h}_{ie}\ln(1-\epsilon_{EV})+\sigma_e^2\beta_i^{(\kappa)}(\br_i,\bw_i)+
\bar{h}_{ie}\sum_{j\neq i}^M \lambda_{ij}^{(\kappa)}(\br_{i},\bw_j,\bw_i),
\end{equation}
which is a concave function satisfying
\[
g_{i,o}(\wk,\rik)= g_{i,o}^{(\kappa)}(\wk,\rik).
\]
Also, following \cite{Shetal17},  the second term in the objective (\ref{sec5a}) is upper bounded
by  the linear function
\begin{equation}\label{dc6}
a_i^{(\kappa)}(\br_i)=\ln(1+\rik)-\ds\frac{\rik}{\rik+1}+\frac{\br_i}{\rik+1},
\end{equation}
while  the first term in (\ref{sec5a}) is lower bounded by  $f_i^{(\kappa)}(\pmb{w})$ defined by (\ref{dc3})
{\color{black} over the trust region (\ref{trustw})}.

We solve the following convex program at the
$\kappa$-th iteration to generate the next feasible point $(w^{(\kappa+1)}, r_u^{(\kappa+1)})$:
\begin{subequations}\label{sec5k}
\begin{eqnarray}
\max_{\pmb{w},\br}\min_{i=1,...,M} [f_i^{(\kappa)}(\bw) -a_i^{(\kappa)}(\br_i)]\quad\mbox{s.t}\quad
{\color{black}(\ref{sec4b}), (\ref{trustw})  (\ref{sec5c}), (\ref{trust1})}, \label{sec5ka}\\
g_{i,o}^{(\kappa)}(\bw,\br_i)\geq 0, \ i=1,...,M. \label{sec5kb}
\end{eqnarray}
\end{subequations}
Then, $r_i^{(\kappa+1)}$ is found from solving the nonlinear equation
\begin{eqnarray}\label{bi1}
\psi_i(\br_i)\triangleq g_{i,o}(\wk,\br_i)=0,\ \ i=1,...,M
\end{eqnarray}
by bisection on $[0,r_{u,i}^{(\kappa+1)}]$ with tolerance $\epsilon_b$ such that
\begin{equation}\label{bi2}
0\leq \psi_i(r_i^{(\kappa+1)})\leq \epsilon_b.
\end{equation}
A bisection on $[r_{l}, r_u]$
for solving $\psi_i(\mathbf{r}_i)=0$ where $\psi_i$ increases in $\mathbf{r_i}>0$ is implemented as
follows:
\begin{itemize}
\item Define $r_i=(r_l+r_u)/2$.  Reset  $r_l=r_i$
if $\psi_i(r_i)<0$. Otherwise reset $r_u=r_i$.
\item Terminate until $0\leq \psi_i(r_i)\leq \epsilon_b$.
\end{itemize}
Like Algorithm \ref{alg1}, Algorithm \ref{alg2} at least converges to a locally optimal solution of (\ref{sec5}).
\begin{algorithm}
\caption{Path-following algorithm for maximin secrecy throughput optimization } \label{alg2}
\begin{algorithmic}
\STATE \textbf{Initialization}: Set $\kappa=0$. Choose an initial feasible point
$(w^{(0)}, r^{(0)})$ for (\ref{sec5}) and calculate $R_{\min}^{(0)}$ as the value of the objective function in
(\ref{sec5}) at $(w^{(0)}, r^{(0)})$.
\REPEAT
\STATE $\bullet$ Solve the
convex optimization problem \eqref{sec5k} to obtain the
solution $(w^{(\kappa+1)}, r_u^{(\kappa+1)})$.
\STATE $\bullet$ Solve the nonlinear equations (\ref{bi1}) to obtain the roots $r_i^{(\kappa+1)}$.
\STATE $\bullet$ Calculate  $R_{\min}^{(\kappa+1)}$ as the value of the objective function in (\ref{sec5}) at
$(w^{(\kappa+1)}, r^{(\kappa+1)})$.
\STATE $\bullet$ Reset
$\kappa+1\rightarrow \kappa$.
 \UNTIL{$\frac{R_{\min}^{(\kappa+1)}-R_{\min}^{(\kappa)})}{R_{\min}^{(\kappa)}} \leq
 \epsilon_{\rm tol}$}.
\end{algorithmic}
\end{algorithm}
Next, the SEE maximization problem (\ref{esec4}) can be formulated as
\begin{subequations}\label{see1}
\begin{eqnarray}
\max_{\pmb{w},\br} \Theta(\pmb{w},\br)\triangleq \sum_{i=1}^M[\ln(1+\ds\frac{|h^H_{ii}\bw_i|^2}{\sum_{j\neq i}|h^H_{ji}\bw_j|^2+\sigma_i^2})-\ln(1+\br_i)]/\pi(\bw)\label{see1a}\\
\mbox{s.t}\quad (\ref{sec4b}), (\ref{sec5b}), (\ref{sec5c}),\label{see1b}\\
\ln(1+\ds\frac{|h^H_{ii}\bw_i|^2}{\sum_{j\neq i}|h^H_{ji}\bw_j|^2+\sigma_i^2})-\ln(1+\br_i)\geq c_i, i=1,\dots, M,
 \label{see1c}
\end{eqnarray}
\end{subequations}
where like (\ref{sec4b}), $c_i$ in (\ref{see1c}) set the QoS threshold for user $i$.

As such, (\ref{see1}) is addressed  by the following iterations with the convergence guaranteed.
\begin{itemize}
\item {\bf Initialization.} Use Algorithm \ref{alg2} to obtain a feasible point $(w^{(0)}, r^{(0)})$ and define
\[
\theta^{(0)}\triangleq \sum_{i=1}^M[f_i(w^{(0)})-\ln(1+r_i^{(0)})]/\pi(w^{(0)}).
\]
\item {\bf $\kappa$-th iteration.} Let $(w^{(\kappa)}, r^{(\kappa)})$ be a feasible point found from
the $(\kappa-1)$th iteration. Define
\[
\theta^{(\kappa)}\triangleq \sum_{i=1}^M[f_i(w^{(\kappa)})-\ln(1+r_i^{(\kappa)})]/\pi(w^{(\kappa)})
\]
and then solve the following convex optimization problem to generate the next feasible point
$(w^{(\kappa+1)}, r_u^{(\kappa+1)})$:
\begin{eqnarray}\label{eseck2}
\ds\max_{\pmb{w},\br}\ \sum_{i=1}^{M}[f_i^{(\kappa)}(\bw) -a_i^{(\kappa)}(\br_i)]
-\Theta(\wk,r^{(\kappa)})\pi(\bw)
\quad\mbox{s.t.}\quad (\ref{sec4b}), (\ref{trustw})  (\ref{sec5c}), (\ref{trust1}), (\ref{sec5kb}),\nonumber\\
f_i^{(\kappa)}(\bw) -a_i^{(\kappa)}(\br_i)\geq c_i, i=1,\dots M.
\end{eqnarray}
Further, $r_i^{(\kappa+1)}$ is found from solving (\ref{bi1}).
\end{itemize}
\section{Robust optimization to compensate users' outate probability }
Now assume that the wiretapped channel $h_{je}$ is in form (\ref{eve}), so the
wiretapped throughput $g_i(\bw)$ for user $i$ at the eavesdropper  is defined via  (\ref{eveas1}) but
\begin{equation}\label{rob}
h_{ji}=\bar{h}_{ji}+\delta \chi_{ji}
\end{equation}
 for $\chi_{ji}\in{\cal CN}(0,I)$ and
$0<\delta<<1$.
The term $\delta\chi_{ji}$ thus represents the channel error in channel state estimation.
Then the user $i$'s throughput $f_i(\bw)$ is implicitly defined through the outage probability as
\begin{equation}\label{userrate3}
\varphi_{i,o}(\bw)\triangleq \ds\max\left\{\ \ln(1+\bR_i)\ :\ \Prob\left(\frac{|\bar{h}_{ii}^H\bw_i|^2}
{\delta |\chi_{ii}^H\bw_i|^2+\sum_{j\neq i}^M|(\bar{h}_{ji}+\delta \chi_{ji})^H\bw_j|^2+\sigma_i^2}<\bR_i\right)<\epsilon
\right\}
\end{equation}
for $\epsilon>0$.

Note that \cite{RTKN14}
\[
\begin{array}{lll}
|(\bar{h}_{ji}+\delta \chi_{ji})^H\bw_j|^2&=&|((1-\delta)\bar{h}_{ji}/(1-\delta)+\delta \chi_{ji})^H\bw_j|^2\\
&\leq& (1-\delta)^{-1}|\bar{h}_{ji}\bw_j|^2+\delta|\chi_{ji}^H\bw_j|^2,
\end{array}
\]
which implies
\[
\begin{array}{lll}
\ds\frac{|\bar{h}_{ii}^H\bw_i|^2}
{\delta |\chi_{ii}^H\bw_i|^2+\sum_{j\neq i}^M|(\bar{h}_{ji}+\delta \chi_{ji})^H\bw_j|^2+\sigma_i^2}&\geq&\\
\ds\frac{|\bar{h}_{ii}^H\bw_i|^2}
{(1-\delta)^{-1}\sum_{j\neq i}^M|\bar{h}_{ji}^H\bw_j|^2+\delta\sum_{j=1}^M|\chi_{ji}^H\bw_j|^2+\sigma_i^2}.&&
\end{array}
\]
Consequently,
\begin{eqnarray}
\ds\Prob\left(\frac{|\bar{h}_{ii}^H\bw_i|^2}
{\delta |\chi_{ii}^H\bw_i|^2+\sum_{j\neq i}^M|(\bar{h}_{ji}+\delta \chi_{ji})^H\bw_j|^2+\sigma_i^2}<\bR_i\right)&\leq&\nonumber\\
\ds \Prob\left(\frac{|\bar{h}_{ii}^H\bw_i|^2}
{(1-\delta)^{-1}\sum_{j\neq i}^M|\bar{h}_{ji}^H\bw_j|^2+\delta\sum_{j=1}^M|\chi_{ji}^H\bw_j|^2+\sigma_i^2}<\bR_i\right).&&
\label{eva1}
\end{eqnarray}
\begin{mypro}\label{evarate} It is true that
\begin{eqnarray}
\varphi_{i,o}(\bw)&\geq&\bar{\varphi}_{i,o}(\bw)\nonumber\\
&\triangleq&\ds\max\{\ \ln(1+\bR_i)\ :\nonumber\\
&&\ds \Prob\left(\frac{|\bar{h}_{ii}^H\bw_i|^2}
{(1-\delta)^{-1}\sum_{j\neq i}^M|\bar{h}_{ji}^H\bw_j|^2+\delta\sum_{j=1}^M|\chi_{ji}^H\bw_j|^2+\sigma_i^2}<\bR_i\right)<\epsilon\}.\label{userrate4}
\end{eqnarray}
\end{mypro}
\Prf  By (\ref{eva1}), if $\bR_i>0$ such that
\[
\Prob\left(\frac{|\bar{h}_{ii}^H\bw_i|^2}
{(1-\delta)^{-1}\sum_{j\neq i}^M|\bar{h}_{ji}^H\bw_j|^2+\delta\sum_{j=1}^M|\chi_{ji}^H\bw_j|^2+\sigma_i^2}<\bR_i\right)
<\epsilon
\]
then
\[
\Prob\left(\frac{|\bar{h}_{ii}^H\bw_i|^2}
{\delta |\chi_{ii}^H\bw_i|^2+\sum_{j\neq i}^M|(\bar{h}_{ji}+\delta \chi_{ji})^H\bw_j|^2+\sigma_i^2}<\bR_i\right)<\epsilon
\]
and (\ref{userrate4}) follows. \qed

Applying (\ref{side12}) in Appendix I for
\[
a=|\bar{h}_{ii}^H\bw_i|^2, b=(1-\delta)^{-1}\sum_{j\neq i}^M|\bar{h}_{ji}^H\bw_j|^2+\sigma_i^2
\]
gives
\begin{eqnarray}
\varphi_{i,o}(\bw)\geq\ds\max\ \{\ \ln(1+\bR_i)\ :\ \delta\left[\delta_M||\bw_{\min}||^2+
\frac{M-1}{2}||\bw_{\min}||^2\ln\frac{\varphi_i(\bw,\bR_i)}{||\bw_{\min}||^2}\right]\leq\nonumber\\
 \varphi_i(\bw,\bR_i)\},\label{r1}
\end{eqnarray}
where
\[
\begin{array}{rll}
\varphi_i(\bw,\bR_i)&\triangleq& \ds\frac{|\bar{h}_{ii}^H\bw_i|^2}{\bR_i}-\left[(1-\delta)^{-1}\sum_{j\neq i}^M|\bar{h}_{ji}^H\bw_j|^2+\sigma_i^2\right],\\
0<\delta_M&\triangleq& \ds -\left(\ln\epsilon
-\ln M +\frac{1}{M}\sum_{i=1}^M\ln\Gamma(i)+\frac{M-1}{2}\ln\delta\right)\\
&=&\ln\epsilon^{-1}
+\ln M -\ds\frac{1}{M}\sum_{i=1}^M\ln\Gamma(i)+\frac{M-1}{2}\ln\delta^{-1},
\end{array}
\]
and
\[
||\bw_{\min}||^2=\min_{i=1,\dots,M}||\bw_i||^2.
\]
Recall that $\Gamma(i)$ are defined from (\ref{side12}).

Therefore, the problem of secrecy rate maximin optimization (\ref{sec4}) is formulated by
\begin{subequations}\label{r2}
\begin{eqnarray}
\max_{\pmb{w},\bR,\br}\min_{i=1,...,M} [\ln(1+\bR_i)-\ln(1+\br_i)]\quad\mbox{s.t}\quad (\ref{sec4b}), (\ref{sec5b}), (\ref{sec5c}),\label{r2a}\\
\varphi_i(\bw,\bR_i)>0, i=1,\dots, M,\label{r2b}\\
\delta\ds\left[\delta_M||\bw_{\min}||^2
+\frac{M-1}{2}||\bw_{\min}||^2\ln\frac{\varphi_i(\bw,\bR_i)}{||\bw_{\min}||^2}\right]\leq \varphi_i(\bw,\bR_i), i=1,\dots, M,\label{r2c}
\end{eqnarray}
\end{subequations}
where $\ln(1+\bR_i)-\ln(1+\br_i)$ in (\ref{r2a}) represents a lower bound for the user $i$'s secrecy throughput.

Constraints (\ref{sec5b}), (\ref{r2b})-(\ref{r2c}) in (\ref{r2}) are nonconvex, which need to be innerly approximated at
each iteration. Let $(w^{(\kappa)}, R^{(\kappa)}, r^{(\kappa)})$ be a feasible point for (\ref{r2}) found from the $(\kappa-1)$th iteration. We have provided an inner approximation for (\ref{sec5b}) by (\ref{trust1}) and (\ref{sec5kb}).
Note that $|\bar{h}_{ii}^H\bw_i|^2/\bR_i$ is a convex function, so
\[
|\bar{h}_{ii}^H\bw_i|^2/\bR_i\geq \ell^{(\kappa)}_i(\bw_i,\bR_i)
\]
for
\[
\ell^{(\kappa)}_i(\bw_i,\bR_i)
\triangleq  2\Re\{(\wik )^H\bar{h}_{ii}\bar{h}_{ii}^H\bw_i\}/R_i^{(\kappa)}
-\bR_i|\bar{h}_{ii}^H\wik |^2/(R_i^{(\kappa)})^2,
\]
which is the linearization of  $|\bar{h}_{ii}^H\bw_i|^2/\bR_i$ at $(\wik , R_i^{(\kappa)})$. Therefore,
the nonconvex constraint (\ref{r2b}) is innerly approximated by the convex constraint
\begin{equation}\label{r2be}
\ell^{(\kappa)}_i(\bw_i,\bR_i)>(1-\delta)^{-1}\sum_{j\neq i}^M|\bar{h}_{ji}^H\bw_j|^2+\sigma_i^2, i=1,\dots, M.
\end{equation}
Furthermore, for
\[
x_i^{(\kappa)}\triangleq \frac{\varphi_i(w^{(\kappa)},R_i^{(\kappa)})}{||\wmink||^2}
\]
it is true that
\[
\ln\frac{\varphi_i(\bw,\bR_i)}{||\bw_{\min}||^2}\leq \ln(x_i^{(\kappa)})-1 +\frac{\varphi_i(\bw,\bR_i)}{x_i^{(\kappa)}||\bw_{\min}||^2}
\]
that yields
\[
||\bw_{\min}||^2\ln\frac{\varphi_i(\bw,\bR_i)}{||\bw_{\min}||^2}\leq \left(\ln(x_i^{(\kappa)})-1\right)||\bw_{\min}||^2+
\frac{\varphi_i(\bw,\bR_i)}{x_i^{(\kappa)}}
\]
Constraint (\ref{r2c}) is thus innerly approximated by
\begin{eqnarray}\label{r2ca1}
\delta\left[\ds\frac{M-1}{2}\left(\ln(x_i^{(\kappa)})-1\right)+\delta_M\right]||\bw_{\min}||^2\leq
\left(1-\delta(M-1)/2x_i^{(\kappa)}\right)\varphi_i(\bw,\bR_i),\\
 i=1,\dots, M.\nonumber
\end{eqnarray}
Set
\[
i_{\min}=\mbox{arg}\min_{i=1,\dots, M}||w_i^{(\kappa)}||^2,
\]
i.e.
\[
||w_{i_{\min}}^{(\kappa)}||^2=\min_{i=1,\dots, M}||w_i^{(\kappa)}||^2.
\]
Verifying numerically that $\frac{M-1}{2}\left(\ln(x_i^{(\kappa)})-1\right)+\delta_M\geq 0$ and $1-\delta(M-1)/2x_i^{(\kappa)}\geq 0$, we use
\begin{equation}\label{ap1}
\varphi_i(\bw,\bR_i)\geq \varphi_i^{(\kappa)}(\bw,\bR_i)\triangleq \ell^{(\kappa)}_i(\bw_i,\bR_i)-(1-\delta)^{-1}\sum_{j\neq i}^M|\bar{h}_{ji}^H\bw_j|^2-\sigma_i^2
\end{equation}
in providing the following convex inner approximation of (\ref{r2ca1}) for each $i=1,\dots, M$:
\begin{eqnarray}\label{r2ca2+}
\delta\left[\ds\frac{M-1}{2}\left(\ln(x_i^{(\kappa)})-1\right)+\delta_M\right]||\bw_{i_{\min}}||^2
\leq \left(1-\delta(M-1)/2x_i^{(\kappa)}\right)\varphi_i^{(\kappa)}(\bw,\bR_i),\\
i=1,\dots, M.\nonumber
\end{eqnarray}
Accordingly, the next feasible point $(w^{(\kappa+1)},R_l^{(\kappa+1)}, r_u^{(\kappa+1)})$ is generated at
the $\kappa$-th iteration by the optimal solution of the convex optimization problem
\begin{equation}\label{sec8A}
\max_{\pmb{w},\bR,\br}\min_{i=1,...,M} [A_i^{(\kappa)}(\bR_i)-a_i^{(\kappa)}(\br_i)]\quad\mbox{s.t}\quad
(\ref{sec4b}),  (\ref{sec5c}), (\ref{trust1}), (\ref{sec5kb}),  (\ref{r2be}), (\ref{r2ca2+}).
\end{equation}
At the same $\kappa$-th iteration, $r_i^{(\kappa+1)}$ is found from solving (\ref{bi1})
by bisection on $[0,r_{u,i}^{(\kappa+1)}]$ such that (\ref{bi2}),
while $R_i^{(\kappa+1)}$ is found from solving
\begin{equation}\label{nonlin1}
\zeta_i(\bR_i)=0
\end{equation}
by bisection on a segment
\begin{equation}\label{nonlinear1B}
\left[R_{l,i}, R_{u,i} \right]
\end{equation}
such that
\begin{equation}\label{bi3}
-\epsilon_b\leq \zeta_i(R_i^{(\kappa+1)})\leq 0.
\end{equation}
 for
\begin{equation}\label{nonlinear1A}
\zeta_i(\bR_i)\triangleq -\varphi_i(w^{(\kappa+1)},\bR_i)+\delta\frac{M-1}{2}||w_{\min}^{(\kappa+1)}||^2\ln\frac{\varphi_i(w^{(\kappa+1)},
\bR_i)}{||w_{\min}^{(\kappa+1)}||^2}+ \delta\delta_M||w_{\min}^{(\kappa+1)}||^2.
\end{equation}
Both $R_{l,i}$ and $R_{u,i}$ in (\ref{nonlinear1B}) can be easily determined
as follow. If $\zeta_i(R_l^{(\kappa+1)})>0$ set $R_{u,i}=R_l^{(\kappa+1)}$ and
$R_{l,i}=R_l^{(\kappa+1)}/\nu$ with the smallest integer $\nu$ such that $\zeta_i(R_l^{(\kappa+1)}/\nu)<0$.
Otherwise, $\zeta_i(R_l^{(\kappa+1)})<0$ set $R_{l,i}=R_l^{(\kappa+1)}$ and
$R_{u}^{(\kappa+1)}=\nu R_l^{(\kappa+1)}$ with the smallest integer $\nu$ such that
$\zeta_i(\nu R_l^{(\kappa+1)})>0$.

\begin{algorithm}
\caption{Path-following algorithm for maximin secrecy throughput optimization } \label{alg4}
\begin{algorithmic}
\STATE \textbf{Initialization}: Set $\kappa=0$. Choose an initial feasible point
$(w^{(0)}, R^{(0)}, r^{(0)})$ for (\ref{r2}) and calculate $R_{\min}^{(0)}$ as the value of the objective function in
(\ref{r2}) at $(w^{(0)}, R^{(0)}, r^{(0)})$.
\REPEAT
\STATE $\bullet$ Solve the
convex optimization problem \eqref{sec8A}  to obtain the
solution $(w^{(\kappa+1)}, R_{l}^{(\kappa+1)}, r_u^{(\kappa+1)})$.
\STATE $\bullet$ Solve the nonlinear equations (\ref{bi1}) to obtain the roots $r_i^{(\kappa)}$.
\STATE $\bullet$ Solve the nonlinear equations (\ref{nonlin1}) for $\zeta_i(\bR_i)$ defined by (\ref{nonlinear1A})
to obtain the roots $R_i^{(\kappa+1)}$.
\STATE $\bullet$ Calculate  $R_{\min}^{(\kappa+1)}$ as the value of the objective function in (\ref{r2}) at
$(w^{(\kappa)}, R^{(\kappa+1)}, r^{(\kappa+1)})$.
 \STATE $\bullet$ Reset
$\kappa+1\rightarrow \kappa$.
 \UNTIL{$\frac{R_{\min}^{(\kappa+1)}-R_{\min}^{(\kappa+1)})}{R_{\min}^{(\kappa)}} \leq
 \epsilon_{\rm tol}$}.
\end{algorithmic}
\end{algorithm}
An initial  feasible $(w^{(0)},R^{(0)}, r^{(0)})$ can be easily found as follows:
taking $w^{(0)}$ and $r^{(0)}$ as the optimal solution of (\ref{sec5}) and $R_i^{(0)}$ is found from
solving $-\epsilon_b\leq \zeta_i(\bR_i)\leq 0$ for
\[
\zeta_i(\bR_i)\triangleq -\varphi_i(w^{(0)},\bR_i)+\delta\frac{M-1}{2}||w_{\min}^{(0)}||^2\ln\frac{\varphi_i(w^{(0)},
\bR_i)}{||w_{\min}^{(0)}||^2}+ \delta\delta_M||w_{\min}^{(0)}||^2
\]
by bisection on $[R_{l,i}, R_{u,i}]$. Here
\[
R_{u,i}=\frac{|h^H_{ie}w_i^{(0)}|^2}{\sum_{j\neq i}
|h^H_{je}w_j^{(0)}|^2+\sigma_e^2}
\]
while $R_{l,i}=R_{u,i}/\nu$ with the smallest integer such that
 $\zeta_i(R_{u,i}/n)<0$.

Next, we address the EE maximization (\ref{esec4}) by the following iterations with the convergence guaranteed.
\begin{itemize}
\item {\bf Initialization.} Use Algorithm \ref{alg4} to obtain a feasible point $(w^{(0)}, R^{(0)}, r^{(0)})$ and define
\[
\theta^{(0)}\triangleq \sum_{i=1}^M[\ln(1+R_i^{(0)})-\ln(1+r_i^{(0)})]/\pi(w^{(0)}).
\]
\item {\bf $\kappa$-th iteration.} Let $(w^{(\kappa)}, R^{(\kappa)}, r^{(\kappa)})$ be a feasible point found from
the $(\kappa-1)$th iteration.
Define
\[
\theta^{(\kappa)}\triangleq \sum_{i=1}^M[\ln(1+R_i^{(\kappa)})-\ln(1+r_i^{(\kappa)})]/\pi(w^{(\kappa)})
\]
and then solve  the two following convex optimization problems to generate the next feasible point
$(w^{(\kappa+1)}, R_l^{(\kappa+1)}, r_u^{(\kappa+1)})$:
\begin{eqnarray}\label{eseck4}
\ds\max_{\pmb{w},\br}\ \sum_{i=1}^{M}[A_i^{(\kappa)}(\bR_i) -a_i^{(\kappa)}(\br_i)]
-\theta^{(\kappa)}\pi(\bw)\nonumber\\
\quad\mbox{s.t.}\quad (\ref{sec4b}),  (\ref{sec5c}), (\ref{trust1}), (\ref{sec5kb}),  (\ref{r2be}), (\ref{r2ca2+}),\nonumber\\
A_i^{(\kappa)}(\bR_i) -a_i^{(\kappa)}(\br_i)\geq c_i, i=1,\dots, M,
\end{eqnarray}
Further, $r_i^{(\kappa+1)}$ is found from solving (\ref{bi1}), while $R_i^{(\kappa+1)}$ is found from solving
(\ref{nonlinear1B}) till satisfactory of (\ref{bi3}) for $\zeta(\bR_i)$ defined by (\ref{nonlinear1A}).
\end{itemize}


\section{Simulation}
This section presents numerical results to demonstrate  the efficiency of the proposed algorithms.
Each transmitter is equipped with $N_t=4$ antennas. Scenarios of $M\in\{2,5,6\}$ pairs with the noise variance $\sigma_i^2=\sigma_e^2=1$ mW are simulated. All entries of channels $h_{je}$ and $h_{ie}$ in (\ref{sec1}) and (\ref{sec2})
are generated by independent and identically distributed complex normal random variables of zero mean and unit variance.
 The drain efficiency $1/\zeta$ of power amplifier in (\ref{esec4}) is
$40 \%$ with the circuit power of each transmit antenna $P_a=1.25$ mW.
The computation tolerance for terminating all proposed Algorithms is $\epsilon_{\rm tol}=10^{-4}$.
The obtained information throughput results are divided by $\ln(2)$ for expressing secrecy throughputs
in  bps/Hz and secure energy efficiencies in bits/J/Hz.

In the below discussion, the terms ``Perfect CSI'', ``EV outage'', and ``User outage '' correspond to the scenarios discussed in Sections III, Section IV with the eavesdropper outage probability $\epsilon_{EV}\in\{0.1, 0.6\}$ in (\ref{eveas1}), and
section V with the channel  error bound $\delta=0.001$ in (\ref{rob}) and user outage probability $\epsilon_c=0.1$ in (\ref{userrate3}), respectively.

\vspace*{-0.3cm}
\subsection{Maximin secrecy throughput optimization}
\begin{figure}
\centering
\includegraphics[width=4in]{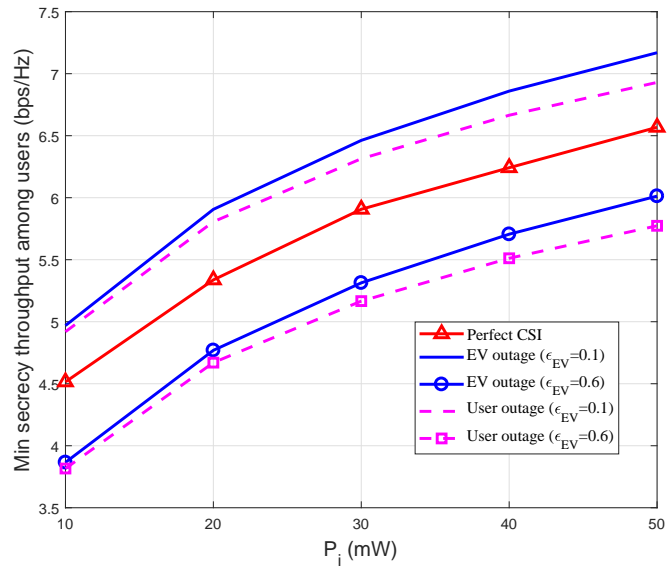}
\caption{Minimal secrecy throughput among users versus the transmit power limitation $P_i$ with $M=2$.}
\label{fig:Maxmin_rate_M2}
\end{figure}

\begin{figure}
\centering
\includegraphics[width=4in]{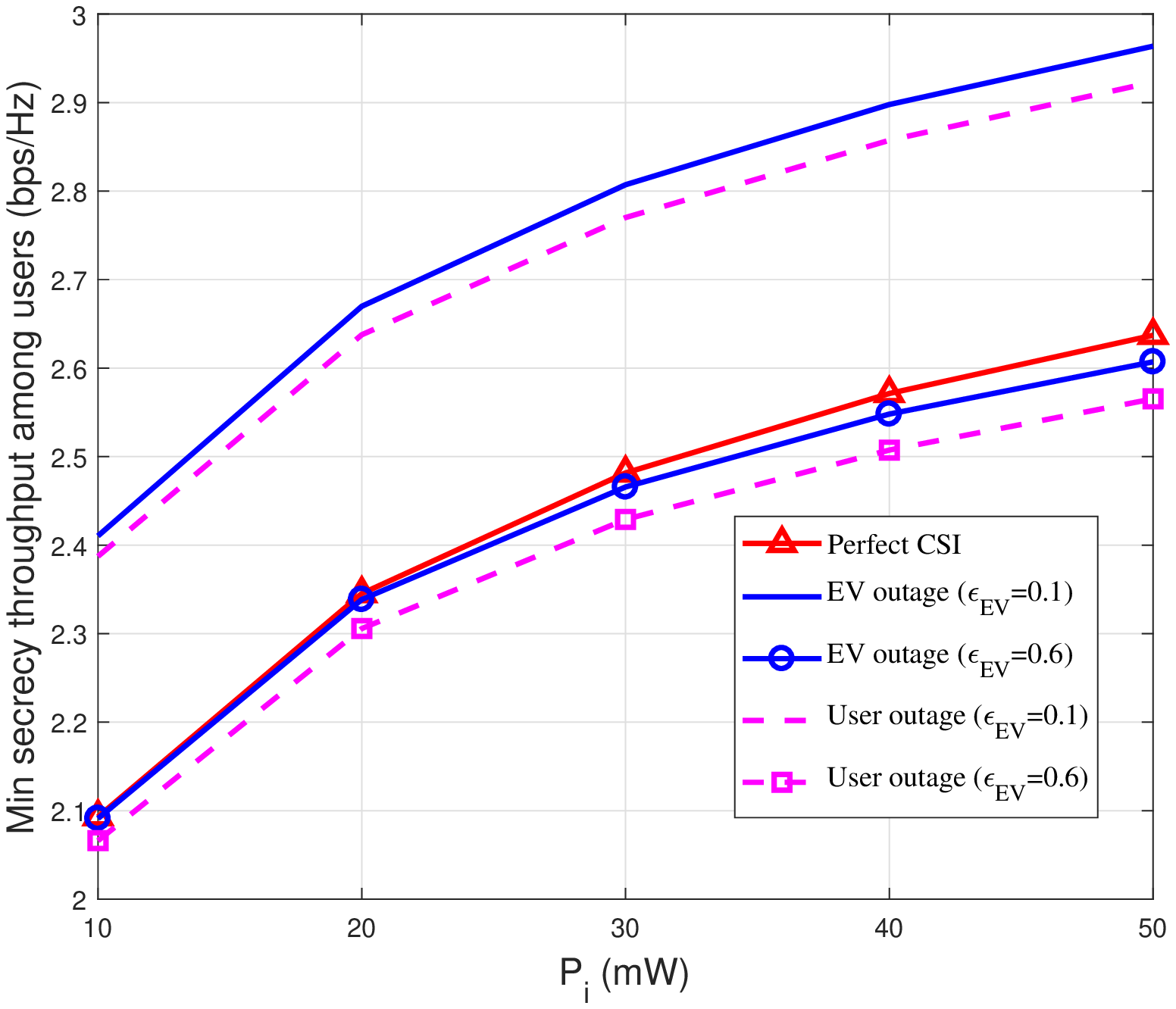}
\caption{Minimal secrecy throughput among users versus the transmit power limitation $P_i$ with $M=5$.}
\label{fig:Maxmin_rate_M5}
\end{figure}

\begin{figure}
\centering
\includegraphics[width=4in]{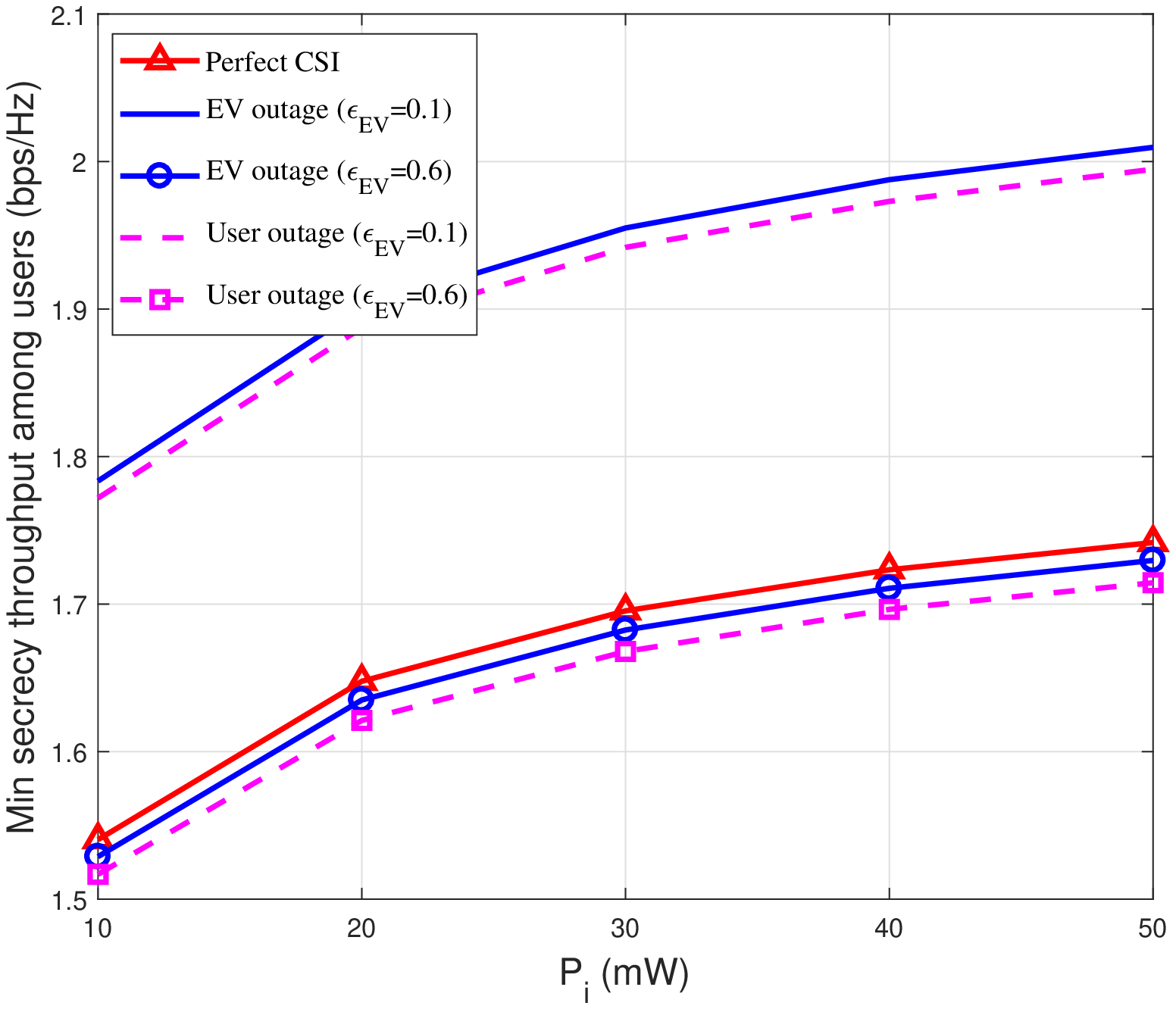}
\caption{Minimal secrecy throughput among users versus the transmit power limitation $P_i$ with $M=6$.}
\label{fig:Maxmin_rate_M6}
\end{figure}
This subsection analyzes the impact of channel uncertainties to the users' achievable secrecy throughput.
Figs. \ref{fig:Maxmin_rate_M2}, \ref{fig:Maxmin_rate_M5} and \ref{fig:Maxmin_rate_M6} plot the users'
minimum secrecy throughput versus the transmit power limitation $P_i$ varying from 10 mW to 50 mW for $M=2$, $M=5$ and $M=6$, respectively. Intuitively, the secrecy throughput increases in the transmitted power limitation $P_i$. In each case
of $M$, both ``EV outage'' and ``User outage'' with the small outage probability  $\epsilon_{EV}=0.1$
achieve better secrecy throughputs than ``Perfect CSI", but the latter achieves
better secrecy throughputs than the formers with the large outage probability $\epsilon_{EV}=0.6$.
This outcome is not surprised  because the instantaneous wiretapped throughput defined by (\ref{eveas})
is actually higher than the throughput outage defined by (\ref{eveas1}) at small outage probabilities $\epsilon_{EV}$.
These figures also show that the secrecy output performance is deteriorated with  the increased
number of transmitter-user pairs, which leads to a stronger  inter-user interference hurting the users' throughput.

Table \ref{tab:Maxmin_rate} provides the average number of iterations
required to solve the problem of  maximin secrecy throughput optimization
 for the above three cases with $M=2$, $M=5$ and $M=6$, respectively.
On average, the proposed algorithms converge in less than $10$, $20$ and $22$ iterations, for $M=2$, $M=5$ and $M=6$, respectively.

\begin{table}
   \centering
   \caption{Average number of iterations for maximin secrecy throughput optimization with $M\in\{2,5,6\}$.}
   \begin{tabular}{ | c | c | c | c | c | c | }
    \hline
   $P_i$ (mW) & 10  &  20 &  30 &  40 &  50  \\
   \hline
   Perfect CSI & 9/12/13 & 8/15/17 & 10/16/16 & 9/18/19 & 8/18/20 \\
    \hline
   EV outage ($\epsilon_{EV}=0.1$) & 5/12/14 & 7/15/17 & 6/17/17 & 7/18/18 & 6/17/20 \\
    \hline
   User outage ($\epsilon_{EV}=0.1$) & 5/8/8 & 3/9/10 & 4/7/12 & 5/11/12 & 4/10/11 \\
    \hline
   EV outage ($\epsilon_{EV}=0.6$) & 8/14/14 & 8/17/18 & 7/17/20 & 8/19/20 & 6/20/22 \\
    \hline
  User outage ($\epsilon_{EV}=0.6$) & 6/9/7 & 6/12/12 & 3/11/10 & 4/12/15 & 4/13/14 \\
    \hline
\end{tabular}
\label{tab:Maxmin_rate}
\end{table}

\vspace*{-0.3cm}
\subsection{Secure energy efficiency maximization}
\begin{figure}
\centering
\includegraphics[width=4in]{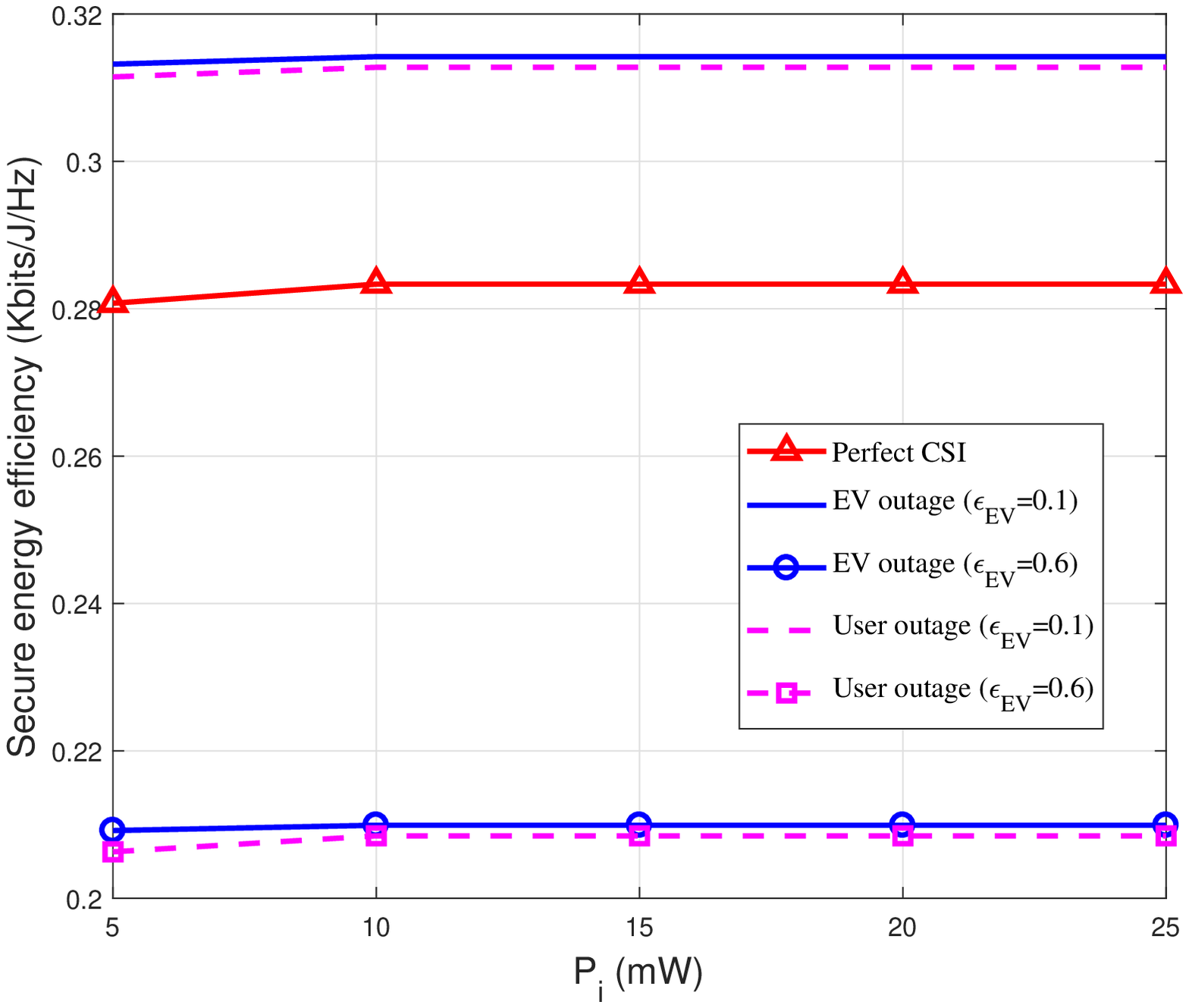}
\caption{Energy efficiency versus the transmit power limitation $P_i$ with $M=2$.}
\label{fig:EE_P_M2}
\end{figure}

\begin{figure}
\centering
\includegraphics[width=4in]{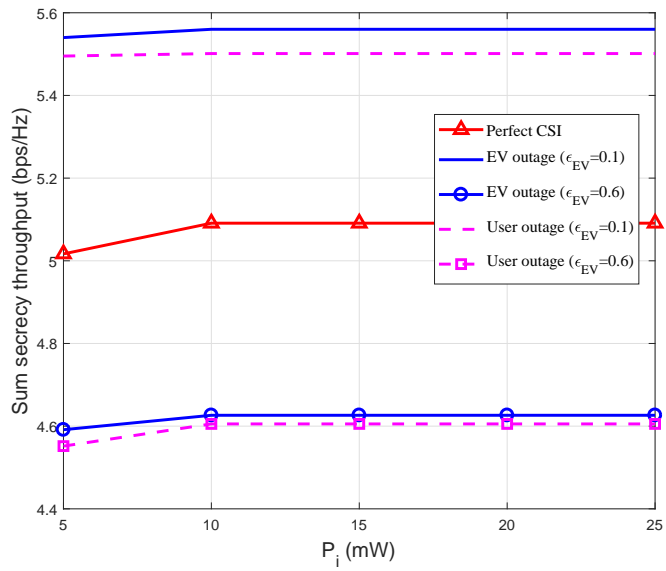}
\caption{Sum throughput versus the transmit power limitation $P_i$ with $M=2$.}
\label{fig:rate_P_M2}
\end{figure}

\begin{figure}
\centering
\includegraphics[width=4in]{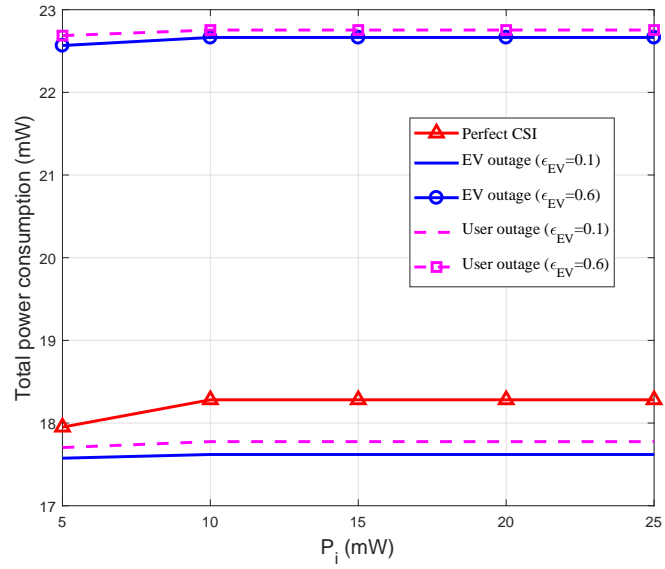}
\caption{Total power consumption versus the transmit power limitation $P_i$ with $M=2$.}
\label{fig:power_P_M2}
\end{figure}

\begin{figure}
\centering
\includegraphics[width=4in]{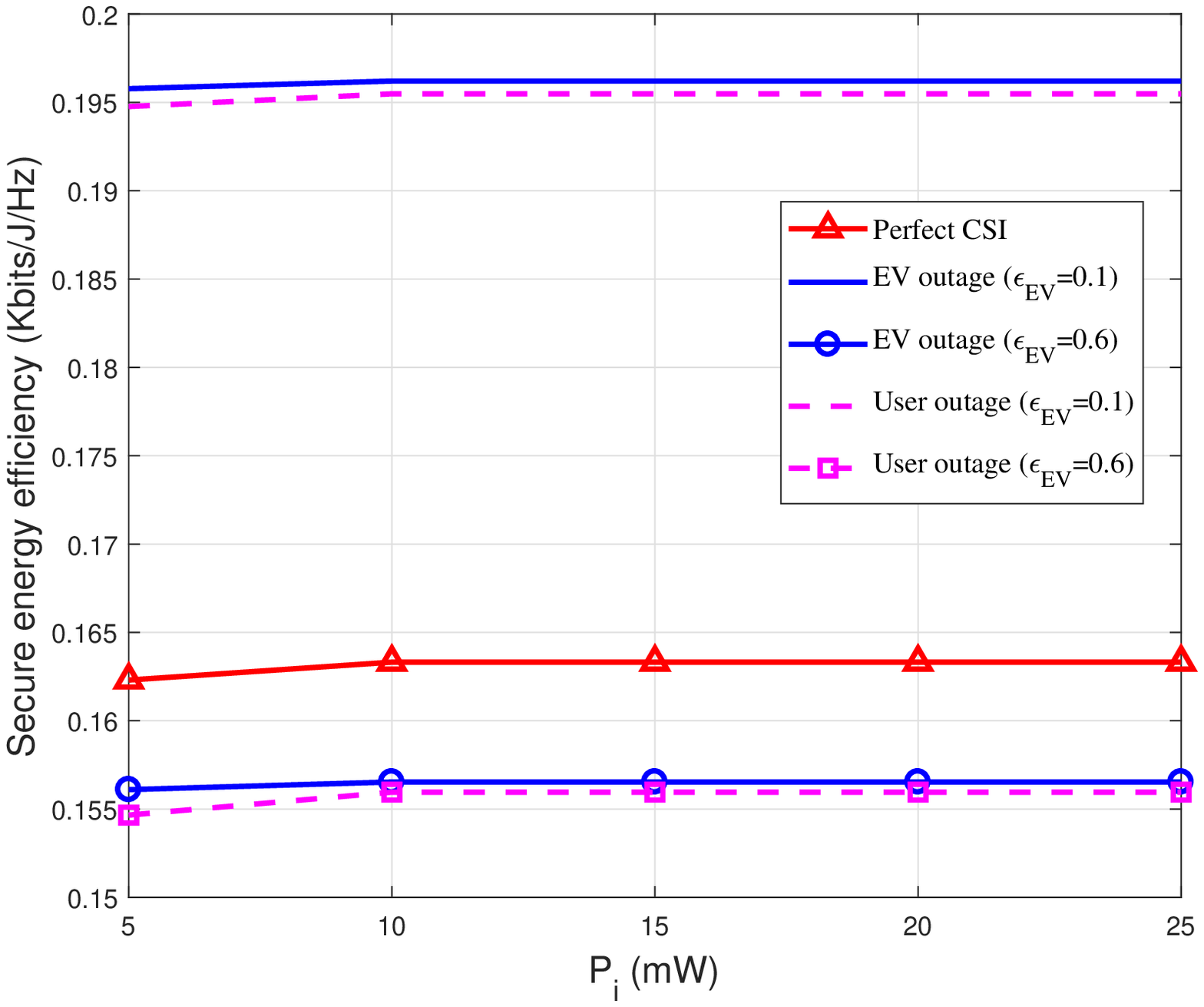}
\caption{Energy efficiency versus the transmit power limitation $P_i$ with $M=5$.}
\label{fig:EE_P_M5}
\end{figure}

\begin{figure}
\centering
\includegraphics[width=4in]{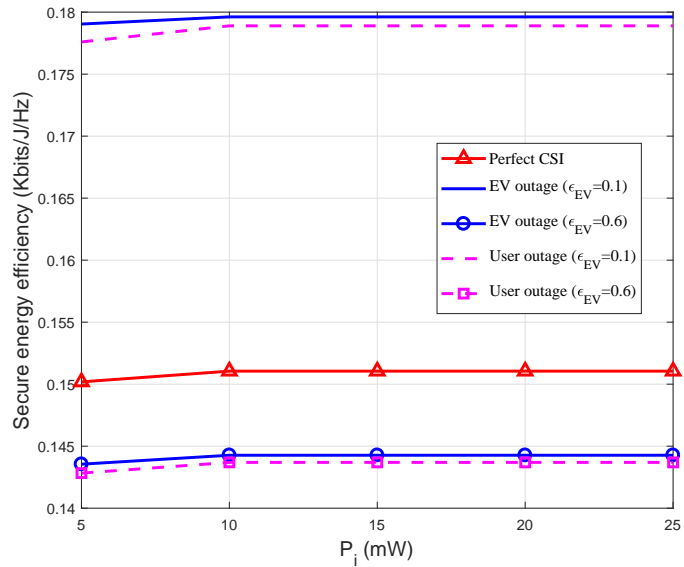}
\caption{Energy efficiency versus the transmit power limitation $P_i$ with $M=6$.}
\label{fig:EE_P_M6}
\end{figure}

This subsection examines the performance of the proposed
SEE maximization algorithms. The threshold
$c_i$ in (\ref{esec4b}) for QoS is 2 bps/Hz, 1 bps/Hz and 0.6 bps/Hz for $M=2$, $M=5$ and $M=6$, respectively. The transmit power limitation $P_i$ varies from $5$ mW to $25$ mW.
Fig. \ref{fig:EE_P_M2} shows that ``EV outage'' with the small outage probability
$\epsilon_{EV}=0.1$ significantly outperforms other cases. The corresponding sum secrecy throughput and
total transmit power plotted in Figs. \ref{fig:rate_P_M2} and \ref{fig:power_P_M2} particularly explain this.
``EV outage'' with $\epsilon_{EV}=0.1$ achieves higher sum secrecy throughput in Fig. \ref{fig:rate_P_M2}
and consumes less power in Fig. \ref{fig:power_P_M2}.
Furthermore, the SEE performances saturates when the transmit power limitation exceeds the threshold $10$ mW.
In the region of small transmit power limitation, the denominator of SEE is dominated by the circuit power
so the SEE is maximized by maximizing the sum secrecy throughput in the numerator.
However, in larger regions of transmit power limitation, the denominator of SEE
becomes to be dominated by the actual transmit power, which by Fig. \ref{fig:power_P_M2}
saturates after $P_i=10$ mW, making the sum secrecy throughput and SEE behave similarly
in Figs. \ref{fig:EE_P_M2} and \ref{fig:rate_P_M2}. Further, by
Fig. \ref{fig:EE_P_M5} and \ref{fig:EE_P_M6}, SEE follows a similar pattern for $M=5$ and $M=6$, respectively.

Lastly, the average number of iterations is provided by
Table \ref{tab:EE_P}, which particularly shows that our proposed SEE maximization algorithm on average converges in less than 16, 24 and 28 iterations for M = 2, M = 5 and M = 6, respectively.

\begin{table}
   \centering
   \caption{Average number of iterations for secure energy efficiency maximization with $M\in\{2,5,6\}$.}
   \begin{tabular}{ | c | c | c | c | c | c | }
    \hline
   $P_i$ (mW) & 5  &  10 &  15 &  20 &  25  \\
   \hline
   Perfect CSI & 10/14/19  & 13/18/23  & 14/20/25  & 14/21/27  & 15/22/28 \\
    \hline
   EV outage ($\epsilon_{EV}=0.1$) & 10/14/16  & 12/17/21  & 12/18/21 &  13/19/22 &  14/19/22 \\
    \hline
   User outage ($\epsilon_{EV}=0.1$) & 3/7/9 & 4/7/9 & 5/9/10 & 3/8/9 & 6/10/11 \\
    \hline
   EV outage ($\epsilon_{EV}=0.6$) & 11/16/18 &  13/19/23  & 15/22/25 &  16/24/26  & 16/23/28 \\
    \hline
   User outage ($\epsilon_{EV}=0.6$) & 5/8/11 & 7/8/11 & 4/9/10 & 6/10/12 & 5/10/12  \\
    \hline
\end{tabular}
\label{tab:EE_P}
\end{table}

\section{Conclusions}
For a wireless network of multiple transmitter-user pairs overhear by an eavesdropper, we have considered the
beamforming design to maximize either the users' secrecy throughput or the network's secure energy efficiency under
QoS constraints in terms of users' secrecy throughput thresholds. At different levels of channel knowledge,
we have developed path-following algorithms of low complexity but rapid convergence for computation. The
provided simulations have not only shown the efficiency of the developed algorithms but also linked the outage probability
with the secrecy degree.   Extensions to multi-cell coordinated  beamforming are underway.
\section*{Appendix I: Outage probability inequalities}
We derive bounds for
\begin{eqnarray}
&&\ds\Prob\left(\frac{a}{\delta\ds\sum_{i=1}^M|\la \chi_i,\bw_{i}\ra|^2+b}<r\right)\label{side3}\\
&\Leftrightarrow&\ds\Prob\left(a/r-b<\delta\ds\sum_{i=1}^M|\la \chi_i,\bw_{i}\ra|^2\right)\label{side3e}
\end{eqnarray}
for $a>0$, $b>0$ and $r>0$. Here $\chi_i\in {\cal CN}(0,I)$ while $\bw_i$ are deterministic complex vectors.

Note that
\[
|\la \chi_i,\bw_{i}\ra|^2=||\bw_i||^2|\la \chi_i,\bw_{i}/||\bw_i||\ra|^2=||\bw_i||^2p_i
\]
where $p_i$ is an exponential distribution with the unit mean.

As
\[
\frac{a}{\delta\ds\sum_{i=1}^M|\la \chi_i,\bw_{i}\ra|^2+b}<\frac{a}{b}
\]
the probability in (\ref{side3}) is not zero if and only if
\begin{equation}\label{side3a}
r<a/b \Leftrightarrow a/r-b>0.
\end{equation}
For
\[
||\bw_{\min}||^2\triangleq \ds\min_{i=1,\dots, M}||\bw_{i}||^2.
\]
it follows that
\[
\begin{array}{lll}
\Prob\left(\ds\sum_{i=1}^M|\la \chi_i,\bw_{i}\ra|^2<(a/r-b)/\delta\right)&=&\ds
\Prob\left(\sum_{i=1}^M||\bw_i||^2p_i<(a/r-b)/\delta\right)\\
&\leq&\ds\Prob\left(\sum_{i=1}^M||\bw_{\min}||^2p_i<(a/r-b)/\delta\right)\\
&=&\ds\Prob\left(\sum_{i=1}^Mp_i<(a/r-b)/\delta||\bw_{\min}||^2\right)\\
&=&\ds\int_{\ds\sum_{i=1}^Mt_i< \frac{a-rb}{\delta||\bw_{\min}||^2}}
\prod_{i=1}^Me^{-t_i} dt_1\cdots dt_M.
\end{array}
\]
Using the representation
\[
u(x)=\frac{1}{2\pi}\int_{-\infty}^{\infty}\frac{e^{\ds x(\jmath\omega+\beta)}}{\jmath\omega+\beta}d\omega
\]
for the unit step function \cite{Aletal16} leads to
\allowdisplaybreaks[4]
\begin{eqnarray}
\ds\int_{\ds\sum_{i=1}^Mt_i< \frac{a-rb}{\delta||\bw_{\min}||^2}}
\prod_{i=1}^Me^{-t_i} dt_1\cdots dt_M&=&\nonumber\\[0.2cm]
\ds\frac{1}{2\pi}\int_{0}^{\infty}\cdots\int_{0}^{\infty}\int_{-\infty}^{\infty}\frac{e^{\ds \left[\ds
\frac{a/r-b}{\delta||\bw_{\min}||^2}-\sum_{i=1}^Mt_i\right]\ds(\jmath \omega+\beta)}}{\jmath\omega+\beta}
\left(\prod_{i=1}^Me^{-t_i} dt_i\right)d\omega&=&\nonumber\\[0.2cm]
\ds\frac{1}{2\pi}\int_{0}^{\infty}\cdots\int_{0}^{\infty}\int_{-\infty}^{\infty}
\prod_{i=1}^Me^{\ds  -t_i(1+\jmath\omega+\beta) }dt_1\dots dt_M\frac{e^{\ds (\jmath\omega+\beta)\frac{a/r-b}{\delta||\bw_{\min}||^2}}}{\jmath\omega+\beta}d\omega&=&\nonumber\\[0.3cm]
\ds\frac{1}{2\pi}\int_{-\infty}^{\infty}
\ds\frac{1}{(\ds 1+\jmath\omega+\beta)^{M}} \frac{e^{\ds (\jmath\omega+\beta)\frac{a/r-b}{\delta||\bw_{\min}||^2}}}{\jmath\omega+\beta}d\omega&=&\nonumber\\[0.3cm]
\ds\frac{1}{2\pi}\int_{-\infty}^{\infty}
\ds\left[\frac{1}{\jmath\omega+\beta}-\sum_{i=1}^M\frac{1}
{(\ds 1+\jmath\omega+\beta)^{i}} \right] e^{\ds(\jmath\omega+\beta)\frac{a/r-b}{\delta||\bw_{\min}||^2}}d\omega,&&\label{add5}
\end{eqnarray}
where for the last equality we have used
\begin{equation}\label{ind1}
\frac{1}{x(1+x)^M}=\frac{1}{x}-\sum_{i=1}^M\frac{1}{(1+x)^i},
\end{equation}
which  can be proved by mathematical induction. Indeed, it is obvious that
\[
\frac{1}{x(1+x)}=\frac{1}{x}-\frac{1}{1+ x},
\]
i.e. (\ref{ind1}) holds true for $M=1$. Suppose that (\ref{ind1}) is true for $M=n$,
i.e.
\[
\frac{1}{x(1+ x)^n}=\frac{1}{x}-\sum_{i=1}^n\frac{1}{(1+x)^i}.
\]
Then
\[
\begin{array}{lll}
\ds\frac{1}{x}-\sum_{i=1}^{n+1}\frac{1}{(1+x)^i}&=&\ds
(\frac{1}{x}-\sum_{i=1}^n\frac{1}{(1+x)^i})
-\frac{1}{(1+x)^{n+1}}\\
&=&\ds\frac{1}{x(1+x)^n}-\frac{1}{(1+x)^{n+1}}\\
&=&\ds\frac{1}{x(1+x)^{n+1}},
\end{array}
\]
i.e. (\ref{ind1}) is true for $M=n+1$, completing the proof for (\ref{ind1}).

Furthermore, by \cite[(28)-(29)]{Aletal16}
\[
\begin{array}{c}
\ds\frac{1}{2\pi}\int_{-\infty}^{\infty}\frac{e^{\ds x(\jmath\omega+\beta)}}{\jmath\omega+\beta}d\omega=1\quad\mbox{for}\quad
x>0,\\
\ds\frac{1}{2\pi}\int_{-\infty}^{\infty}\frac{e^{\ds x(\jmath\omega+\beta)}}{(1+\jmath\omega+\beta)^i}d\omega
=\frac{e^{-x}x^{i-1}}{\Gamma(i)}\quad\mbox{for}\quad x>0,
\end{array}
\]
where $\Gamma(i)\triangleq\int_0^{\infty}t^{i-1}/e^tdt$, for which
$\Gamma(1)=\Gamma(2)=1$, $\Gamma(3)=2$, $\Gamma(4)=6$, $\Gamma(5)=24$.

We thus obtain
\begin{eqnarray}
\ds\Prob\left(\sum_{i=1}^M|\la \chi_i,\bw_{i}\ra|^2<(a/r-b)/\delta\right)\leq 1-e^{-(a/r-b)/\delta||\bw_{\min}||^2}\sum_{i=1}^M\frac{(a/r-b)^{i-1}}{\ds\Gamma(i)
 (\delta||\bw_{\min}||^2)^{i-1}},\label{side5}
\end{eqnarray}
or
\begin{eqnarray}
\ds\Prob\left(\sum_{i=1}^M|\la \chi_i,\bw_{i}\ra|^2\geq (a/r-b)/\delta\right)&=&1-\ds\Prob\left(\sum_{i=1}^M|\la \chi_i,\bw_{i}\ra|^2< (a/r-b)/\delta\right)\nonumber\\
&\geq&\ds 1 -\left[ 1-e^{-(a/r-b)/\delta||\bw_{\min}||^2}\sum_{i=1}^M\frac{(a/r-b)^{i-1}}{\ds\Gamma(i)
 (\delta||\bw_{\min}||^2)^{i-1}} \right]\nonumber\\
&=& e^{-(a/r-b)/\delta||\bw_{\min}||^2}\sum_{i=1}^M\frac{(a/r-b)^{i-1}}{\ds\Gamma(i)
 (\delta||\bw_{\min}||^2)^{i-1}}.\label{side5e}
\end{eqnarray}
Analogously,
\begin{eqnarray}
\ds\Prob\left(\sum_{i=1}^M|\la \chi_i,\bw_{i}\ra|^2<(a/r-b)/\delta\right)\geq
 1-e^{\ds -(a/r-b)/\delta||\bw_{\max}||^2}\sum_{i=1}^M\frac{(a/r-b)^{i-1}}{\Gamma(i)(\delta||\bw_{\max}||^2)^{i-1}},\label{side6}
\end{eqnarray}
for
\[
||\bw_{\max}||^2\triangleq \ds\max_{i=1,\dots, M}||\bw_{i}||^2.
\]
Therefore, we arrive at the following result.
\begin{myth}\label{twosideth2} The following two-sided inequalities hold true:
\begin{eqnarray}
\ds e^{\ds \ds-(a/r-b)/\ds\delta||\bw_{\min}||^2}
\ds\sum_{i=1}^M\frac{(a/r-b)^{i-1}}{\ds\Gamma(i)(\delta||\bw_{\min}||^2)^{i-1}}
&\leq&\label{side7}\\[0.2cm]
\ds\Prob\left(a/r-b<\delta\sum_{i=1}^M|\la \chi_i,\bw_{i}\ra|^2\right)=\Prob\left(\frac{a}{\delta\ds\sum_{i=1}^M|\la \chi_i,\bw_{i}\ra|^2+b}\leq r\right)&\leq&\nonumber\\[0.2cm]
\ds e^{\ds -(a/r-b)/\delta||\bw_{\max}||^2}\sum_{i=1}^M\frac{(a/r-b))^{i-1}}{\Gamma(i)(\delta||\bw_{\max}||^2)^{i-1}}.&&\label{side8}
\end{eqnarray}
\qed
\end{myth}
Now, by Cauchy inequality
\[
\begin{array}{lll}
\ds\sum_{i=1}^M\frac{(a/r-b)^{i-1}}{\ds\Gamma(i)(\delta||\bw_{\min}||^2)^{i-1}}&\geq&\ds
M\left(\prod_{i=1}^M\frac{(a/r-b)^{i-1}}{\ds\Gamma(i)(\delta||\bw_{\min}||^2)^{i-1}} \right)^{1/M} \\
&=&\ds\frac{M}{(\prod_{i=1}^M\Gamma(i))^{1/M}}\left(\frac{(a/r-b)}{\delta||\bw_{\min}||^2}\right)^{(M-1)/2}.
\end{array}
\]
Therefore, it follows from (\ref{side7}) that
\begin{eqnarray}\label{side9}
\ds e^{-(a/r-b)/\delta||\bw_{\min}||^2}\ds\frac{M}{(\prod_{i=1}^M\Gamma(i))^{1/M}}\left(\frac{(a/r-b)}{\delta||\bw_{\min}||^2}\right)^{(M-1)/2}
&\leq&\nonumber\\
 \ds\Prob\left(\frac{a}{\delta\ds\sum_{i=1}^M|\la \chi_i,\bw_{i}\ra|^2+b}\leq r\right).&&
\end{eqnarray}
Then
\begin{eqnarray}
\ds\max\ \left\{r\ :\  \Prob\left(\frac{a}{\delta\ds\sum_{i=1}^M|\la \chi_i,\bw_{i}\ra|^2+b}\leq r\right)\leq \epsilon\right\}&\geq&\nonumber\\
\ds\max\ \left\{r\ :\ e^{-(a/r-b)/\delta||\bw_{\min}||^2}\ds\frac{M}{(\prod_{i=1}^M\Gamma(i))^{1/M}}\left(\frac{(a/r-b)}{\delta||\bw_{\min}||^2}\right)^{(M-1)/2}
\leq \epsilon\right\}&=&\label{side10}\\
\ds\max\ \left\{r\ :\ -\frac{a/r-b}{\delta||\bw_{\min}||^2}+(\ln M -\frac{1}{M}\sum_{i=1}^M\ln\Gamma(i))\right.&&\nonumber\\
+\ds\left.\frac{M-1}{2}\left(\ln(a/r-b)-\ln\delta-\ln||\bw_{\min}||^2\right)\leq \ln\epsilon\right\}&=&\label{side11}\\
\ds\max\ \left\{r\ :\ -\frac{a/r-b}{||\bw_{\min}||^2}+\delta(\ln M -\frac{1}{M}\sum_{i=1}^M\ln\Gamma(i))\right.&&\nonumber\\
+\ds\left.\delta\frac{M-1}{2}\left(\ln(a/r-b)-\ln\delta-\ln||\bw_{\min}||^2\right)\leq \delta\ln\epsilon\right\}&=&\nonumber\\
\ds\max\ \left\{r\ :\ -\frac{a/r-b}{||\bw_{\min}||^2}+\delta\frac{M-1}{2}\left(\ln(a/r-b)-\ln||\bw_{\min}||^2\right)\leq\right.&&\nonumber\\
\ds\left. \delta\left(\ln\epsilon
-\ln M +\frac{1}{M}\sum_{i=1}^M\ln\Gamma(i)+\frac{M-1}{2}\ln\delta\right)\right\}.&&
\label{side12}
\end{eqnarray}
Note that for $M=1$  it follows from (\ref{side7}) and (\ref{side8}) that
\[
\Prob\left(\frac{a}{|\la \chi,\bw\ra|^2+b}<r\right)=e^{-(a/r-b)/||\bw||^2},
\]
which is a known result since $|\la \chi,\bw\ra|^2$ is an exponential distribution with mean $||\bw||^2$:
\[
\begin{array}{lll}
\ds\Prob\left(\frac{a}{|\la \chi,\bw\ra|^2+b}<r\right)&=&\ds\Prob\left(a/r-b< \la \chi,\bw\ra|^2\right)\\
&=&\ds\int_{a/r-b}^{\infty}e^{\ds -t/||\bw||^2}/||\bw||^2dt\\
&=&e^{-(a/r-b)/||\bw||^2}.
\end{array}
\]
Particularly,
\begin{equation}\label{par1}
\max\{ r:\ \Prob\left(\frac{a}{\delta|\la \chi,\bw\ra|^2+b}<r\right)\leq \epsilon\}=\frac{a}{b+\delta||\bw||^2\ln\epsilon^{-1}}.
\end{equation}
\section*{Appendix II: Bernstein-type inequality and its conservativeness}
There is an approach, which  is based on the following Bernstein-type inequality \cite{Be09} of rough estimation.
\begin{myth}\label{btheorem}\cite[Lemma 0.2]{Be09} Suppose that $A$ is a symmetric matrix and $z$ is Gaussian with zero mean
and identity covariance. Then
\begin{equation}\label{be1}
\text{Prob}\left(z^HAz\geq \mbox{trace}(A)+2||A||\sqrt{x}+2\lambda^+_{\max}(A)x\right)\leq \exp(-x),
\end{equation}
where $\lambda^+_{\max}(A)=\max\{\lambda_{\max},0\}$.
\end{myth}
One can use inequality (\ref{be1}) for an inner approximation of the set
\begin{equation}\label{be2}
R_{\epsilon}(a,b)\triangleq \left\{r\ :\
\text{Prob}\left(\delta\sum_{i=1}^M|\la \chi_i,\bw_i\ra|^2>\ds\frac{a}{r}-b\right)\leq \epsilon\right\}.
\end{equation}
By setting $z=[\chi_i^T]^T_{i=1,\dots,M}$ and
\[
A=\mbox{diag}[\bw_i\bw_i^H]_{i=1,\dots,M}
\]
we have  $\sum_{i=1}^M|\la \chi_i,\bw_i\ra|^2=z^HAz$ and
$\mbox{trace}(A)=\sum_{i=1}^M||\bw_i||^2$, and
\[
\lambda_{\max}(A)=\max_{i=1,\dots,M}[||\bw_i||^2]\quad\mbox{and}\quad
||A||=\sqrt{\sum_{i=1}^M||\bw_i||^4}.
\]
According to (\ref{be1})
\[
\text{Prob}\left(z^HAz\geq \mbox{trace}(A)+2||A||\sqrt{\ln \epsilon^{-1}}+2\lambda_{\max}(A)\ln\epsilon^{-1}\right)\leq \epsilon.
\]
Therefore,  $r\in R_{\epsilon}(a,b)$ if
\begin{eqnarray}
&&\ds(\frac{a}{r}-b)/\delta\geq \mbox{trace}(A)+2||A||\sqrt{\ln \epsilon^{-1}}+2\lambda_{\max}(A)\ln\epsilon^{-1}\nonumber\\[0.2cm]
&\Leftrightarrow&\ds(\frac{a}{r}-b)/\delta\geq \sum_{i=1}^M||\bw_i||^2+
2\sqrt{\sum_{i=1}^M||\bw_i||^4}\sqrt{\ln \epsilon^{-1}}+\ds 2\max_{i=1,\dots,M}[||\bw_i||^2]\ln\epsilon^{-1}\nonumber\\[0.2cm]
&\Leftrightarrow&\ds r\leq a/\left[b+\delta\left(\sum_{i=1}^M||\bw_i||^2+
2\sqrt{\sum_{i=1}^M||\bw_i||^4}\sqrt{\ln \epsilon^{-1}}
+\ds 2\max_{i=1,\dots,M}[||\bw_i||^2]\ln\epsilon^{-1}\right)\right]
\label{be3}
\end{eqnarray}
which is too conservative compared with (\ref{side12}). For instance, for $M=1$, (\ref{be3}) means
\[
r=a/(b+\delta||\bw||^2(1+2\sqrt{\ln\epsilon^{-1}}+2\ln\epsilon^{-1}))
\]
which is very conservative compared with (\ref{par1}).
\section*{Appendix III: Basic deterministic inequalities}
For every $x>0$, $y>0$, $\bar{x}>0$ and $\bar{y}>0$,
\begin{eqnarray}
\ds\ln(1+1/xy)&\geq&\ds\ln(1+1/\bar{x}\bar{y})+\ds\frac{1/\bar{x}\bar{y}}{1+1/\bar{x}\bar{y}}(2-x/\bar{x}-
y/\bar{y}),
\label{sec5.2}
\end{eqnarray}
which follows from the convexity of function $\ln(1+1/xy)$ in the domain $\{x>0, y>0\}$. Furthermore,
\begin{equation}\label{conv1}
\ds\ln (1+x/y)\leq \ln(1+\bar{x}/\bar{y})+\ds\frac{1}{1+\bar{x}/\bar{y}}
(x/y-\bar{x}/\bar{y}),
\end{equation}
which follows from the concavity of function $\ln(1+z)$ in the domain $\{z>0\}$.
Lastly, based on the inequality
\begin{equation}\label{conv2}
x^2/t\geq 2 (\bar{x}/\bar{t})x-(\bar{x}^2/\bar{t}^2)t\quad\forall\ x>0, \bar{x}>0, t>0, \bar{t}>0
\end{equation}
that follows from the convexity of $x^2/t$, we have the following inequality
\begin{equation}\label{conv3}
\frac{r}{||\bw||^2}\geq 2(\sqrt{\bar{r}}/||\bar{w}||^2)\sqrt{r}-(\bar{r}/||\bar{w}||^4)||\bw||^2\quad
\forall\ r>0, \bar{r}>0, \bw\in\mathbb{C}^N, \bar{w}\in\mathbb{C}^N.
\end{equation}
\bibliographystyle{ieeetr}
\bibliography{outage}
\end{document}